\documentstyle[prl,aps,multicol]{revtex}
\input{epsf}
\begin{document}
\draft
\title{Quantum Chaos and the Black Body Radiation}                      

\author{Giulio Casati}
\address{International Center for the Study of dynamical Systems, Universit\`a degli Studi dell' Insubria,  
Via Valleggio,11,
22100 Como, Italy}
\address{Istituto Nazionale di Fisica della Materia, 
Unit\`a di Milano, Via Celoria 16, 20133 Milano, Italy}
\address{Istituto Nazionale di Fisica Nucleare, Sezione di Milano,
Via Celoria 16, 20133 Milano, Italy} 

\date{\today}
\maketitle
\begin{abstract}
We discuss a mechanical model which mimics the main features of the radiation matter interaction in the
 black body problem. 
The pure classical dynamical evolution, with a simple discretization of
the action variables, leads
 to the Stefan- Boltzmann law and to the Planck distribution 
without any additional statistical assumption.
\end {abstract}

\begin{multicols}{2}
\narrowtext

The problem of black body radiation occupies a central role in the history of physics. The long debate  around its 
properties stimulated a profound revision of old and firmly established concepts and eventually led to the birth of
 quantum mechanics. Yet it is somehow unfortunate that such a profound revision of a fundamental theory has been 
based on a problem of so great complexity. Indeed, from the dynamical viewpoint, one is dealing with a nonlinear 
system with infinite degrees of freedom. Needless to say, such an infinite set of nonlinear differential equations 
is not solvable. What is more important, and perhaps less noticed, is that even the statistical description do not 
rest on solid grounds. As a matter of facts, classical ergodic theory which is now quite well developed, is valid 
for systems with a {\it finite} number $N$ of degrees of freedoms. A main difficulty stems from the fact that the 
two limits $N \rightarrow \infty$ and $ t\rightarrow\infty$ do not commute. The quantum statistical description  
is even more complex since it involves the additional difficulty that also 
the two limits
$t\rightarrow \infty$, $\hbar\to 0$ do not commute. 

In view of the relevance of the black body problem, it is perhaps worthwhile, 
after hundred years, a re-examination 
of the statistical properties with the help of modern computers and at 
the light of the recent progress in the study 
of nonlinear classical and quantum dynamical systems.  This implies to tackle 
the problem 
of the statistical behaviour of infinite systems.

A distinctive feature of the radiation matter interaction in the black body
is that each normal mode, or field's oscillator, interacts with the matter's 
degrees of freedom only, and that the strength of this interaction
decreases as the mode's frequency increases. In this paper we introduce a simplified model which however shares the 
main features of the black body.
We consider a system of $N$ oscillators with mass $m_i = c/i^2$ and 
frequency $\omega_i= \sqrt{k/m_i} = \alpha {i}$ with $\alpha=\sqrt{k/c}$.
When at its central 
position $x=0$ each oscillator collides elastically with a particle of mass 
$ M>>m_i$. Therefore the whole system is conservative with total energy 

\begin{eqnarray} 
\label{model}
 E= E_0 +\sum_{i=1}^{N}  E_i = (1/2)M{V^2}+ \sum_{i=1}^{N} I_i\omega_i + h.c.
\end{eqnarray}

where $E_0$ is the energy of the heavy particle and $I_i$ are the actions of 
the oscillators.
The interaction between oscillators and the heavy particle is provided by 
the hardcore(h.c.) collisions.
In our numerical computations we have taken the mass of the heavy particle
$M= (\sqrt{5}+1)/2$, $c=0.51$, $k=0.1$ so that $\omega_i=\alpha {i}$
with $\alpha \approx 0.443$.
Notice that our system is of a billiard type and therefore the trajectory
does not depend on the total energy $E$ which is merely a
scale factor.
We recall that the one-dimensional system of two hard core point particles 
with fixed boundary conditions is equivalent 
to the billiard in a right triangle which is assumed to be ergodic and weakly 
mixing\cite{prosen}. In our model the particle exchanges energy with each
oscillator and moreover, after each collision, we assign at random, the
sign of the velocity of the heavy particle. Our classical model has
therefore a high degree of chaoticity. 
We would like to stress that for the purpose of the present paper we only need a mechanisms which conserves the total energy and allows energy exchange between the different oscillators. Therefore, we are not interested in the detailed motion of the heavy particle. Instead we need a 
sufficient degree of chaotic behaviour in order to ensure ergodicity of the 
entire system with $N+1$ degrees of freedom\cite{nota}.

The model system (\ref{model}) is a mechanical version of the one dimensional 
black body problem discussed in \cite{benenti} and one
expects a similar statistical behaviour.
Indeed we have numerically computed
the time averaged energies of the oscillators and of the particle.  As 
expected, we have observed that for any $N$, and 
independently of the initial condition,  the system approaches the equipartition state in which the time 
averaged energies of the oscillators and of the particle are all equal (an example is given by the open squares in fig. 1) . 
Therefore, for any finite value of the total energy $E$, with increasing the 
number  $N$ of oscillators the temperature of the system $T=E/N$ will 
decrease down to zero as $N\rightarrow\infty$ . The mechanisms through which such state is approached 
is the one already envisaged by Jeans\cite{jeans}: energy flows from the
matter(our particle) to higher and higher modes of the electromagnetic field(our oscillators) in such a way that
 the time averaged energy in each mode is zero while the total energy remains constant. Therefore the field 
continuously absorbs energy from the matter, energy will move endless to higher and higher modes, and the whole
 system cools down to absolute zero temperature. This would be the behaviour according to classical laws. 
Fortunately however, this is not the case since our world is governed by quantum mechanics. The latter leads 
to the celebrated Planck distribution which, in one dimension, reads:

\begin{eqnarray} 
\label{planck}
 E_i =E(\omega_i) = \hbar{\omega_i}/(\exp{\beta{\hbar}{\omega_i}}-1)
\end{eqnarray}

The temperature $T=1/\beta$ is finite and connected to the total energy $E_f$ 
in the field (in the oscillators in our case) by the relation
$E_f=\sigma{T^2}$ 
which is known as Stefan-Boltzmann  law (in the one dimensional case).

Expression(\ref{planck}) has been derived by statistical methods. A {\it dynamical} 
derivation based on the quantum theory or on the numerical solution of the 
Schrodinger equation without any statistical assumption is still 
lacking. In 
this sense we are asking, for the quantum case, the same question that was 
posed about 50 years ago by Fermi, Pasta and Ulam\cite{fermi} regarding the 
problem of classical equipartition (which is the limit 
of (\ref{planck}) for $\hbar\to 0$). Equipartition was a very well established 
consequence of classical statistical 
mechanics even though it rested on statistical assumptions. Fermi however, wanted to 
derive this property by a direct solution of Newton equations. The result was unexpected and Fermi 
considered it as  one of the most important of his life. Indeed, as we now know, it 
is at the root of the modern field 
of nonlinear classical dynamics and chaos. It is therefore worthwhile to ask now a similar 
 question: can we, by direct
 numerical integration of the time dependent Schrodinger equation, derive expression(\ref{planck})?

A convenient guide in this direction can be provided by recent progress in the so-called field of quantum chaos. 
One of the main results in the study of this field is the
quantum suppression 
of classically chaotic diffusion\cite{ford,review}. This phenomenon is
known as quantum dynamical localization since is the dynamical
 analog of Anderson localization which takes place in disordered 
solids\cite{fishman}.
Localization represents a strong deviation from quantum ergodicity and only when localization length is larger than
 the sample size then eigenfunctions are extended, quantum ergodicity takes place and statistical methods can be 
applied. For example, in such situation, statistical properties of energy levels can be described 
by random matrix theory. Therefore, 
if the eigenfunctions are localized, then quantum evolution leads to a 
stationary state which is typically exponentially localized around the 
initially excited state and therefore
strongly depends on the initial condition. On the other 
hand, if eigenfunctions are delocalised then the quantum stationary state is 
close to the corresponding 
classical one. In both cases it is not 
clear according to which 
quantum {\it dynamical} mechanisms Planck distribution sets 
in \cite {sred} .

In order to provide a clear answer to the above question it would be necessary to 
compute the quantum evolution 
for a system with many interacting particles like the system (\ref{model}). This is a too difficult task.

\begin{figure}
\centerline{\epsfxsize=9cm \epsfbox{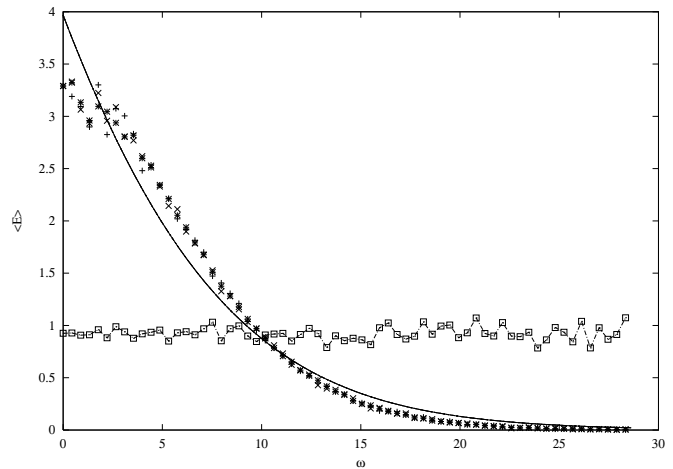}}
\vspace{4mm}
\caption{
Time averaged energy of the particle $E_0$ and of the oscillators
$E_i$, as a function of frequency $\omega$
for the discrete model with $N=64$ and total energy $E=60$.
The averages are taken over a large number $t$ of total
collisions. Initial condition with almost all energy
in the heavy particle: $t=10^7$ (+), $t=4 (10^7)$ (*);classical
model, $t=4 (10^7)$ (open squares). Initial condition with almost
all energy in the oscillator i=32, $t=10^7$(x). The full curve
is the Planck law (2) with $\beta$ given by
eq(3) at $E=60(\beta =0.252)$.
}
\end{figure}

In the following we present instead the results of a numerical integration 
of classical system  (\ref{model}) in which 
however we allow for the actions $I_i$ to take only integers values. Namely, after each collision, which obeys 
the usual classical conservation laws, we substitute the values $I_j=E_j/\omega_j$ with the nearby integers $n_j$. 
The choice between the upper or the lower nearest integer is made at random after each collision. 
The roundoff energy of the oscillator is then given to the heavy particle in order
to ensure total energy conservation.
We will refer 
to this model as to the discrete model to distinguish it from the usual model(\ref{model}) in which classical 
evolution is computed without any approximation. Our surmise, to be verified, is that such a simple discretization 
procedure will qualitatively reproduce the main results of an exact integration of Schrodinger 
equation (with $\hbar=1$). Such possibility was also suggested in \cite{chirikov,berman1,berman2}. 
More recently we have found that application 
of such procedure to the standard map leads to  exponential quantum
localization\cite{casati}.
Clearly the difference between quantum and classical mechanics
goes much beyond the discrete nature of phase space. It is however our hope
to gain a better understanding of their relationship.

In fig.1 we show the time averaged energies of the particle 
and of the oscillators, obtained with the above described numerical
 scheme for $N=64$ and total energy $E=60$.
The thin curve is the
theoretical Planck law (\ref{planck}) with $\hbar=1$ and $\beta$ given
by (\ref{total})with $E=60$.

\begin{eqnarray}
\label{total}
 E = 1/\beta + \sum_{i=1}^{\infty} \omega_i/(\exp{\beta\omega_i}-1)
\end{eqnarray}

 It is quite remarkable that, 
{\it independently on the initial condition}, the discrete model
always reaches the same stationary distribution. We have
also checked that the time-averaged values do not change by 
increasing the integration time. As an example we show in  
fig. 2 the time-averaged energies for few oscillators as a
function of time measured in number of collisions, up to $t=10^7$.
These values do not change by further increasing the integration time.

We will not enter here in 
several questions of detail and simply notice that the 
overall agreement between the numerical results and the theoretical
curve (\ref{planck}) (with the correct value of $\hbar =1$)is sufficiently
satisfactory.
Solution of the classical model starting with the same initial conditions 
leads to equipartition as is shown by the open squares in fig.1.

\begin{figure}
\centerline{\epsfxsize=9cm \epsfbox{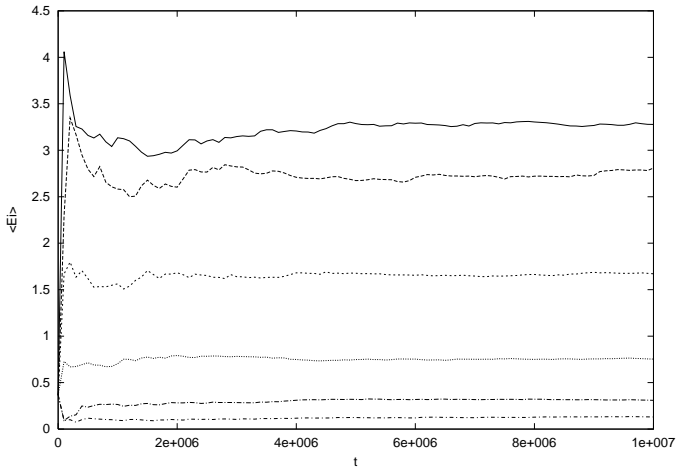}}
\vspace{4mm}
\caption{
Time-averaged energy $<E_i>$ of oscillators as a function of total number
$t$ of
collisions. The curves refer to oscillators (starting from above):
$i=1$, $i=8$, $i=16$, $i=24$, $i=32$, $i=40$. }
\end{figure}

In order to analyse the behaviour of the system
as a function of the energy, at fixed $N$, we plot in fig.3
the total energy $E$ of the system over the
time averaged energy$<E_0>$ of the particle. Clearly in our discret 
model
the classical limit is reached for $E\rightarrow \infty$, since
the effects of discretization become less and less important
as the total energy is increased. Therefore,
for sufficiently large energy, one expects
equipartition which is given by the relation $<E>=(N+1)<E_0>$.
According to standard statistical mechanics, the temperature
is defined as the average kinetic energy $T= <E_0)>$.

\begin{figure}
\centerline{\epsfxsize=9cm \epsfbox{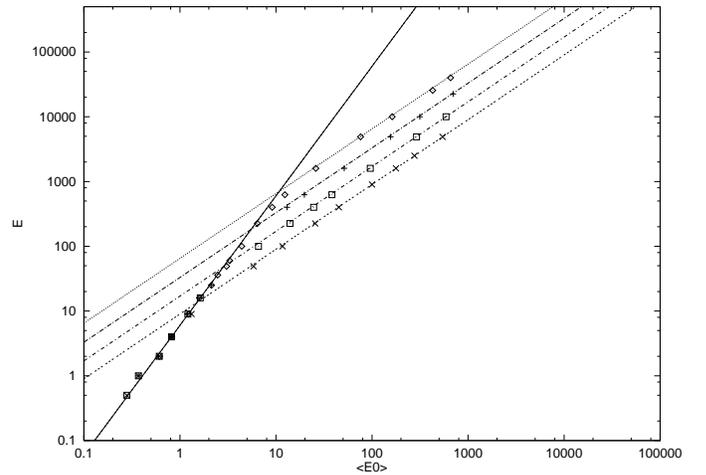}}
\vspace{4mm}
\caption{
The total energy $E$ of the system over the average energy
$<E_0> =T$ of the particle for different $N$ values.
$N=64$(open rombus), $N=32$(+), $N=16$(open squares), $N=8$(x). 
The solid
line has
slope two while the four parallel lines have slope one.
}
\end{figure}

As we decrease energy, at fixed particle number $N$, 
we expect deviation from equipartition. 
It is really remarkable that integration of our discrete
model leads to the Stefan-Bolzmann law which, in our 
one-dimensional case, implies that the total energy increases 
quadratically with the temperature.
 The solid line in fig. 3 is given by $E=\sigma{<E_0>^2}
=\sigma{T^2}$ with $\sigma\approx 6$ (independent on $N$). The four
parallel 
lines are given instead by $E= (N+1)<E_0>$ 
($N=8, 16, 32, 64$)which is the 
exact equipartition law of classical mechanics.
According to our data, the transition from 
Stefan-Boltzmann to equipartition
is quite sharp and
takes place at $E \approx N^2/\sigma$.

\begin{figure}
\centerline{\epsfxsize=9cm \epsfbox{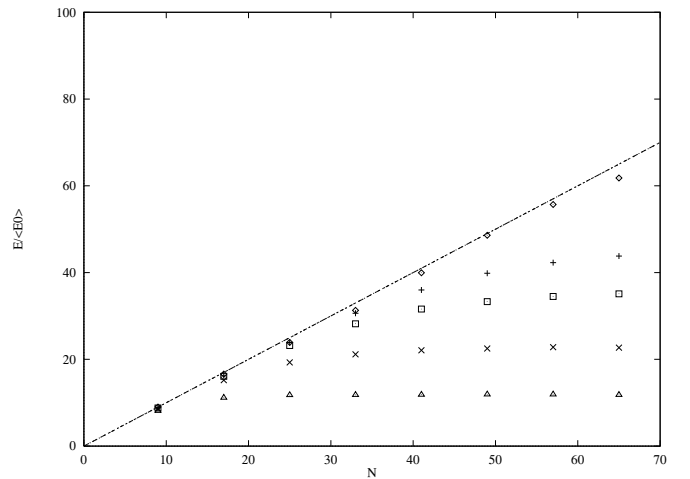}}
\vspace{4mm}
\caption{
Plot of $E/<E_0>$ versus $N$ for different values of total
energy $E$;  $E=25$(open triangles), $E=100$(x), $E=225$(open squares),
$E=400$(+),
$E=1600$(open rombus). The dotted line has slope one and corresponds to equipartition.}
\end{figure}

In order to analyse instead the behaviour of the system
as a function of the number $N$ of oscillators, at 
fixed total energy $E$, we plot in fig.4 the quantity $E/<E_0>$
over  $N$, for
 different values of the total energy $E$. The dotted line
 corresponds to equipartition and it is approached by decreasing $N$, at
fixed energy $E$.
Instead, by increasing $N$ one approaches  a constant 
value  $\propto \sqrt E$. This is again in agreement with
the Stefan-Boltzmann law.
In summary, the results presented in figs 1-4 agree with 
the predictions of quantum mechanics. For a 
given $N$, at low temperature, the Stefan-Boltzmann
law and the Planck distribution are
obtained; as temperature is increased, one
approaches the semiclassical limit in which equipartition 
takes place. 

It may be interesting to examine this transition in relation to the
general theory of quantum dynamical localisation.
According to
\cite{pichard}, the 
scaling {\it Ansatz} is equivalent
to postulating the existence of a function $f(x)$ such that

\begin{eqnarray} 
\label{loc}
 l/N=  f(x),       x= \xi/N
\end{eqnarray}

where $l$ is the actual localization length 
in the sample of finite size $N$,
and $\xi$ is the 
localization length for the infinite sample.
As a measure of localization, following standard
procedure, we take the inverse 
participation ratio:

\begin{eqnarray} 
\label{loc}
 l=<E_f>^2/\sum_{j=1}^{N}<E_j>^2
\end{eqnarray}

\begin{figure}
\centerline{\epsfxsize=9cm \epsfbox{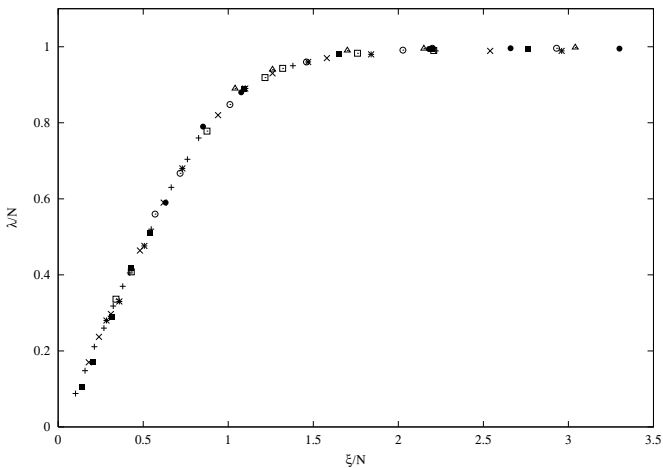}}
\vspace{4mm}
\caption{
The actual localization length $l/N$ for the discrete model
over the localization
length $\xi/N$ for the infinite "sample" for different 
"sample sizes" $N$ and for different energies $E$.
$N=64$(+), $N=56$(x), $N=48$(*), 
$N=40$(open squares), $N=32$(full squares), 
$N=24$(open circles), $N=16$(full circles), $N=8$(open triangles). 
}
\end{figure}

In fig.5 we plot the actual localization length $l/N$ 
versus $\xi/N$ where $\xi$ is computed from 
the distribution 
(\ref{total}). The localisation length $l$ is 
obtained from expression(\ref{loc}) where $<E_i>$
are the time-averaged energies of the oscillators
computed for the discrete model and $<E_f>$ is the
total average energy of the oscillators.
A quite nice scaling behaviour is observed.

Certainly the results presented here require a better understanding. In
particular it is not clear to what extent they are related to the
peculiarity of the radiation matter interaction in which the field modes
interact only through the matter. Needless to say the conclusions we have
drawn here refer to the discrete model (\ref{model}). It is our personal
 opinion that the solution of the time dependent 
Schrodinger equation will lead to the same qualitative 
results of figs. 1-5. This however needs a carefull 
study of true quantum mechanics of system(\ref{model}) or
 of similar models.

\end{multicols}


\begin{thebibliography}{10}
\bibitem{nota} One may consider more realistic models than 
the one studied here. For example one may take the heavy
particle moving between two fixed walls in one dimension 
or in a two-dimensional billiard. However the main results
will remain unchanged.  The main ingredients are the fact 
that the system is chaotic, the total energy is conserved 
and that the oscillators exchange energy only via interaction 
with the heavy particle. The choice of this particular model 
is only due to numerical convenience.
\end{thebibliography}
\end{document}